\documentclass[onecolum,a4,prl, notitlepage]{revtex4-1}
\usepackage{relsize}
\usepackage[countmax]{subfloat}
\usepackage{graphicx}
\usepackage{amssymb, amsmath,amssymb,amsfonts}
\usepackage{amsthm,mathrsfs,amsopn}
\usepackage{dcolumn}
\usepackage{bm}
\usepackage{color}
\usepackage[utf8]{inputenc}
\usepackage[table]{xcolor} 
\usepackage{floatrow}
\usepackage{lettrine}
\usepackage{lineno}
\usepackage{dsfont}
\usepackage[makeroom]{cancel}

\begin{document}

\title{Reply to Comment on “Synchronization dynamics in non-normal networks: the trade-off for optimality”}

\author{Riccardo Muolo$^1$, Timoteo Carletti$^1$, James P. Gleeson$^2$, Malbor Asllani$^3$ \vspace*{.25cm}}
\affiliation{$^1$Department of Mathematics and naXys, Namur Institute for Complex Systems, University of Namur, rue Grafé 2, 5000 Namur, Belgium}
\affiliation{$^2$MACSI, Department of Mathematics and Statistics, University of Limerick, Limerick V94 T9PX, Ireland}
\affiliation{$^3$School of Mathematics and Statistics, University College Dublin, Belfield, Dublin 4, Ireland}

\begin{abstract}
We reply to the recent note ``Comment on Synchronization dynamics in non-normal networks: the trade-off for optimality'' \cite{comment}, showing that the authors base their claims mainly on general theoretical arguments that do not necessarily invalidate the adequacy of our previous study \cite{ourpaper}. In particular, they do not specifically tackle the correctness of our analysis but instead limit their discussion on the interpretation of our results and conclusions, particularly related to the concept of optimality of network structure related to synchronization dynamics. Nevertheless, their idea of optimal networks is strongly biased towards their previous work and does not necessarily correspond to our framework, making their interpretation subjective and not consistent. We bring here further evidence from the existing {and more recent} literature, omitted in the Comment note, that the synchronized state of oscillators coupled through optimal networks, as intended by the authors, can indeed be highly fragile to small but finite perturbations, confirming our original results.
\end{abstract}

\maketitle


\noindent
In the Comment note \cite{comment}, the authors claim we have drawn ambiguous or incorrect conclusions in our recent paper \cite{ourpaper} regarding the lack of optimality of non-normal networks and the failure of the linear stability analysis {as a prediction tool} in such a setting. Here we show that such claims contradict the recent literature (including previous papers written by authors of \cite{comment}) or omit part of the information {already present in our paper. Despite the fact that we found the Comment \cite{comment} harsh and partial, we are grateful to its authors for having publicly expressed their point of view. In fact, such reaction has sparked an immediate interest in the scientific community, in particular the recent studies \cite{sorrentino, fish_bollt} carried out independently by researchers unrelated to the authors of this Reply letter. As we will show in the following, not only do such studies confirm our main results presented in \cite{ourpaper}, in neat contrast with the Comment note \cite{comment}, but they have pushed them further focusing exclusively on the class of optimal networks as intended by the authors of the Comment \cite{comment}.} 
{In the following we will respond to the supposed issues raised in \cite{comment} aiming this way to avoid any possible misunderstanding from the interested readers. However, b}efore we address such claims, we find it necessary to briefly summarize the emergence of real networks' non-normal properties and the consequences in the linear dynamic regime.

A matrix $\textbf{A}$ is defined as non-normal if it does not commute with its self-adjoint: $\textbf{A}\textbf{A}^*\neq \textbf{A}^*\textbf{A}$. Such a property has important consequences for the dynamics of linear systems of the form $\dot{\textbf{x}}=\textbf{A}\textbf{x}$, as well as for nonlinear ones with $\textbf{A}$ the linearized operator. As a prominent example, we mention here the seminal work by Trefethen \textit{et al.} in hydrodynamic stability \cite{trefethen2}. In this context, they used the pseudo-spectrum of $\textbf{J}$ defined as $\sigma_\delta\left(\textbf{J}\right)=\sigma\left(\textbf{J}+\boldsymbol{\Delta}\right)$ for all $||\boldsymbol{\Delta}||\leq \delta$, where $\textbf{J}$ represent the linearised (Jacobian) operator, to determine a lower bound of the transient growth as a function of the non-normality of $\textbf{J}$. Such a relation reads $\sup_{t\geq 0}||e^{\textbf{A}t}||\geq \mathcal{K}$ where we make use of the Kreiss constant $\mathcal{K}=\sup_{\delta \geq 0} \alpha_{\delta}\left(\mathbf{J}\right)\big /\delta$ with $\alpha_{\delta}\left(\mathbf{J}\right)=\sup\Re{\sigma_\delta(\mathbf{J})}$. The latter expression is crucial since it builds a strong link between the pseudo-spectrum and the non-monotonic evolution of the linear dynamics. {Based on that, Trefethen \textit{et al.} have shown that when the amount of non-normality increases beyond some threshold, small but finite perturbations can undergo an initial growth and the nonlinear system can eventually stabilize in another state different from the one near which the perturbations originated.}
As Trefethen $\&$ Embree have often noted in their book \cite{trefethen}, although such outcomes do not contradict the mathematical formulation of the stability theory, there are still neat evidences where linear methods will practically fail to predict the final behavior of the system under investigation. The reason for that is attributed to the fact that an increase in non-normality will drastically reduce the basin of attraction of the linearly stable state, {which we further stress, is defined on local results only.} The latter puts to the fore a general deficiency of local analysis compared to the global one, which is built on the promises that such description can capture the general behavior of the system for initial conditions starting near the stable state. Such results make local stability techniques in a natural or experimental setting, where perturbations will be very small but still finite, of little practical utility to analyze non-normal systems. Recently, a set of systematic statistical studies has brought to light {the fact} that most real networks, including biological, ecological, social, transport, economic, etc. \cite{top_resilience, malbor_teo, me, duccio_stoch, hierarchy}, possess a high level of non-normality with peculiar structural features. {In this regard, it has been recently shown for networked systems in \cite{me} that despite a negative Master Stability Function (MSF) \cite{pecora} indicating a strictly stable fixed point, the orbits starting near such a state could escape from it, stabilizing in other stationary equilibrium.}\\

\noindent\underline{--\textit{{Adequacy of the Linear Stability Analysis (e.g., MSF) and the finite perturbations question}}}\vspace*{.1cm}\\
Based on the considerations above, in our paper \cite{ourpaper}, we have studied synchronization dynamics in non-normal networks based on the MSF formalism. {From the local stability analysis perspective,} we have shown that non-normal networks present spectral features that optimize the stability of the synchronized state. {Still,} the evolution of the nonlinear system evades the {linear prediction because of the reduction of the basin of attraction of the synchronized manifold,} in line with the previous results known in the literature. 
In their original papers \cite{mott_nish1, mott_nish2}, the authors have been limited to a theoretical analysis based exclusively on linear stability analysis and have not performed numerical simulations. Thus they have not {emphasized} the nonlinear behavior and in particular the restriction of the synchronization basin {at variance with the authors of \cite{sorrentino, fish_bollt} who have carried out systematic numerical investigations and consequently confirmed our original findings.}
Furthermore the affirmations made in \cite{comment} are {surprisingly contradictory with} a very recent contribution of the first two authors of the Comment note \cite{mott_nn} agreeing with our statements (pg. 1)

\vspace*{.15cm}

``\textit{Even when a fixed point is linearly stable in a nonlinear system described by ordinary differential equations, if the corresponding Jacobian matrix is non-normal, a small but finite perturbation can transiently grow beyond the validity of the linear approximation and enter into the nonlinear regime, preventing the perturbation from decaying to zero.}''

\vspace*{.15cm}

which makes clear that our results should not be surprising at all regarding the validity limitations of linear approximations {that, as we clearly show in the following, extend also to the case of the linearization about a limit cycle}. Before we address further concerns, we will bring to the attention of the interested reader that similar qualitative results to ours {in} \cite{ourpaper} were independently obtained in \cite{Sun}. We quote {from \cite{Sun}} in the following (pg. 1)

\vspace*{.15cm}

``\textit{Our results show that optimal networks that are more sensitive with respect to structural perturbations (thus having higher sensitivity index) tend to be slower in synchronization, and may sometimes not synchronize at all despite being deemed synchronizable under linear stability analysis.}'' 

\vspace*{.15cm}

Moreover they continue (pg. 7) 

\vspace*{.15cm}

``\textit{This loss of synchronization cannot be interpreted by a linear theory, since the equivalent linear system would always converge to the (global) stable state despite initial transient growth. The actual nonlinear system, however, can have a non-global basin of synchronization. If the initial transient growth is too fast, as seen in some of the very sensitive networks, it “kicks" the state of the system outside of the basin, and thus prevent the system from returning to eventual synchronization.} '' 

\vspace*{.15cm}

{It is worth mentioning that although the sensitive optimal networks have been considered in several works \cite{prl_exp, Fish, PRX}, the authors of \cite{comment} have omitted this fact in their Comment note.} On the other hand, {the observations of \cite{Sun}} do not diminish the novelty of our results since they do not relate to the transient growth behavior that drives such evasion from the linear prediction to the non-normality of such networks.
Furthermore, in our paper {and in full analogy with \cite{trefethen2}}, we have systematically quantified such relationships using techniques such as the pseudo-spectrum (Fig. 5 $a)$), grounded on the strong link between the latter and the transient growth mentioned earlier. 

{As mentioned earlier, the recent discussion triggered by our original paper \cite{ourpaper} has drawn the attention of the scientific community \cite{sorrentino, fish_bollt}. In particular, and similarly to \cite{Sun}{, the authors of such recent studies,} focus exclusively on models of optimal networks following the definition as introduced by the first two authors of the Comment in \cite{mott_nish1, mott_nish2}. Here an excerpt from the conclusions of \cite{sorrentino}}

\vspace*{.15cm}

\textit{{``Our numerical results confirm some of the claims presented in Refs. \cite{ourpaper, footnote} and indicate that networks with non-normal Laplacian matrix are typically characterized by a smaller basin of attraction than networks with a normal Laplacian matrix. Our work also complements some of the early results in \cite{Sun}\dots~  Finally, our work points out the need for an alternative measure of network synchronizability which addresses the important issue of the basin of attraction for the synchronous solution.''}} 

\vspace*{.15cm}

{and the conclusions of \cite{fish_bollt}}

\vspace*{.1cm}

\textit{{``In this work we have examined the stability of the synchronous state of "optimal" non-normal matrices, specifically of non-normal graph Laplacians. For such networks, MSF analysis is no longer useful in some circumstances and we must appeal to other measures to characterize the stability.''}}

\vspace*{.1cm}

{In both cases, the authors agree that \textit{infinitesimal} perturbations make no sense in the real scenarios ``\textit{It is well known that the validity of the MSF approach is limited to infinitesimal perturbations about the synchronous solution (local stability.) However, in all applications, one is interested in the effects of finite perturbations.}'' in \cite{sorrentino} and ``\textit{Of note in real world systems as verified by experiment, noise plays an important role and so finite perturbations cannot be ignored.}" in \cite{fish_bollt}, respectively. Thus, we are confident that the qualitative reproduction of our study \cite{ourpaper} carried out independently by different groups of researchers and unrelated to the authors of this Reply letter are in line with the best scientific practice and should be sufficient to dispel any doubt regarding the validity of our findings.} Similar confirmations can be found also in \cite{Fish}, referring to the use of MSF approach {in non-normal networks} for finite perturbations: ``\textit{In such cases, initial conditions near the synchronization manifold may suffer large transients that can expel trajectories from the expected basin of stability} \cite{Sun, me, malbor_teo}'' \\

\noindent\underline{--\textit{{Optimality of non-normal networks and the averaging method}}}\vspace*{.1cm}\\
Regarding the spectral optimality of the non-normal networks, we have been focused on a network model (a directed chain line with reciprocal links uniformly weighted with $\epsilon$), whose Laplacian always admits a basis of eigenvectors. Such a model converges to the class of optimal networks \cite{mott_nish1, mott_nish2} for $\epsilon\rightarrow 0$, but otherwise, it is generally different from them. In fact, from the Comment note, one could draw the wrong conclusion that our work was focused exclusively on the optimal networks \cite{mott_nish1, mott_nish2}. In contrast, nowhere in our paper have we defined our network model based on such a class of optimal networks. Instead, we have stated that they share similar features related to {strong directedness or Laplacian degeneracy (for $\epsilon=0$), but {they are} otherwise {different} being diagonalizable. Thus our concept of optimality, although based on the same MSF approach, differs from the one of \cite{mott_nish1, mott_nish2} without any prejudice for the latter. On their side, the authors firmly refer to the optimal networks they have previously introduced, but as we have pointed in \cite{ourpaper} different definitions of \textit{optimal networks} already exist in literature \cite{skardal1, skardal2}. Furthermore, there is a striking difference between our model of non-normal networks and the class of optimal networks \cite{mott_nish1, mott_nish2} which are non-diagonalizable, as also correctly pointed out in \cite{sorrentino}. As we have previously noticed in \cite{top_resilience} the geometric degenerancy of the spectrum can bring about a transient growth, as also explained in the Comment note. However, there is a crucial point that {has been} ommitted in {the} Comment \cite{comment}: that such transient growth is possible in a larger class of networks that are not necessarily non-diagonalizable, the non-normal ones. This fact refutes the incorrect assertion made {in} the Comment that we haven't extended the idea of non-normal dynamics to the synchronization phenomenon.} At variance with \cite{mott_nish1, mott_nish2}, we make use of the classical MSF formulation \cite{pecora}. We so require diagonalizable networks with varying imaginary part (tuning the $\epsilon$ parameter) of the Laplacian eigenvalues to quantify the spectral optimality. Such numerical analysis has been presented in Fig. 2 of \cite{ourpaper}{. In our paper, we emphasize that alternatively} analytical insight {can be obtained} by approximating the Master Stability Function by averaging the time-dependent Jacobian over the period $\langle \boldsymbol{\mathcal{J}}\rangle_T$ {(Fig. 3), this approach aims to bridge our} problem of optimality to known results \cite{malbor, malbor2} where the role of the imaginary part of the Laplacian spectrum has been proved analytically. Such an approximation is based on the Magnus expansion, and its validity {for studying the stability of limit cycles is given in \cite{chall} where a detailed analysis of the same model (i.e., Brusselator) has been carried out.} In fact, in our paper, we clearly state that ``\textit{the set of model parameters for which we expect a good agreement with the original model corresponds to the case when higher-order terms are negligible.}''. {This alternative approach is suggested to analytically justify the optimality we obtain numerically in Fig. 2 by varying the $\epsilon$ parameter, but we have not developed such argument further and, contrary to what is stated in the Comment \cite{comment}, our claims of optimality are based on the numerical approach of Fig. 2 rather than the analytical one of Fig. 3. On the other hand, w}e agree that the averaging method does not apply for all models, however because of the latter remark, the counterexample provided in \cite{comment} does not fit in our framework. {R}egarding the use of the averaging method, the authors of the Comment note seem to {contradict previous statements made by some of them}. {In} particular the third author of \cite{comment} has several times advised the averaging of the time-dependent Jacobian matrix as a valid alternative to the MSF \cite{pecora1, pecora2}, for instance (pg. 3 of \cite{pecora1}):

\vspace*{.15cm}

``\textit{We noted that other stability criteria are possible. Each will produce its own master function over the complex coupling plane. Among them are the following four (roughly in order of strictness): (1) Calculate an exponent for an averaged Jacobian [Brown $\&$ Rulkov, 1997a, 1997b],}'' (the second reference, here \cite{rulk}, was already included in our paper when referred to the averaging of the Laplacian). 

\vspace*{.4cm}

\noindent\underline{--\textit{{Further minor points}}}\vspace*{.1cm}\\
{We agree with the authors of the Comment note that we have written ``\textit{(real part of the maximum) Lyapunov exponent}'' which {is redundant} since the Lyapunov exponents are real by definition. Nevertheless, such unfortunate oversight does not compromise the results of our analysis as far as the real part of a real number is still a real number. Also, with ``\textit{autonomous version of the MSF}'' we intended the MSF of an autonomous linearized system as shown in our paper. {We are confident that such unharmful lapses do not compromise our study, and the attentive reader will easily grasp the novelty of our contribution, as was, for instance, the case of the authors of \cite{sorrentino, fish_bollt}}. Also, in the Comment note there is a long discussion about the \textit{bubbling} phenomenon. Although we do agree in principle with {the authors} on it, bringing such a phenomenon as an argument in this discussion can result misleading since non-normality and bubbling are not necessarily connected. What we are stating in our paper is that transient growth and the eventual escaping from the linear stability prediction can be certainly caused by the network non-normality. {We did not observe the bubbling effect in our study, but we are by no means} excluding the possibility that other factors, such as bubbling, can cause similar behavior. Thus we find either in such comparison confusing rather than clarifing.} {As a last note, we did not find neither in the Comment's references, nor in the broader literature, any explicit relation between the non-normality and the robustness of the synchronized regime.}

\vspace*{.1cm}

{To conclude, we want to emphasize that t}here is an increasing research field, among which our paper {sits}, that clearly points towards the same conclusion we reached with our numerical study, i.e., when a non-normal matrix governs the dynamics, a linear stability analysis may fail in predicting the global stability of a system. Let us stress that, in addition, our results have not been tested nor reproduced, making a comparison on the same basis difficult. And last but not least, we want to reemphasize an elementary fact: perturbation theories (where also the MSF method belong), since they were first deployed, were aimed to understand the behavior of nonlinear systems by approximating their dynamical evolution with few terms of series expansions based on a small but still finite parameter and thus the term \emph{infinitesimal}, used by the authors of the Comment, is highly misleading. Such an approximation will be valid to study the linear stability as long as the perturbations lie inside the attraction basin of the state under consideration. Unfortunately, even in the most controlled laboratory conditions, such perturbations cannot be ``infinitesimal,'' and for the case of highly non-normal systems, as is the case of many real networks, the basin of attraction can be drastically small (in our paper, this can be clearly observed in Fig. 5 panels $c_1$ and $c_2)$), thus making linear methods fail to achieve their goal. Finally, for consistency {with} what {is} stated above, we quote our paper where the term ``finite perturbations'' has been abundantly used strongly contrasting the authors' claims in the Comment paper:

\vspace*{.15cm}

``\textit{In a linear regime, a finite perturbation regarding a stable equilibrium goes through a transient amplification (blue curve in Figure 4 $d)$) that is proportional to the level of non-normality before it is eventually reabsorbed in the steady state \cite{trefethen}, while, in the full non-linear system, the finite perturbation could persist indefinitely (red curve in Figure 4 $d)$).}''

\paragraph*{Funding}
R.M. is supported by a FRIA-FNRS PhD fellowship, Grant FC 33443, funded by the Walloon region. The work of J.P.G. and M.A. is partly funded by Science Foundation Ireland (Grants No. 16/IA/4470, No. 16/RC/3918 and No. 12/RC/2289 P2) and co-funded under the European Regional Development Fund. M.A. is also supported by a OBRSS Research Support Scheme no. R23036.

\vspace*{-.5cm}



\begin{thebibliography}{999}
\bibitem{footnote} The authors have cited here the first version of our Reply letter.
\end{thebibliography}
\end{document}